\documentclass[a4paper,twoside,titlepage,12pt]{article}

\usepackage{amssymb,amsmath,amsfonts}
\usepackage{epsfig}

\begin{document}

\begin{titlepage}

\rightline{NBI-HE-00-33}
\rightline{hep-th/0007147}
\rightline{July, 2000}
\vskip 2cm

\centerline{\Large \bf Open Branes in}
\vskip 0.2cm
\centerline{\Large \bf Space-Time Non-Commutative}
\vskip 0.2cm
\centerline{\Large \bf Little String Theory}
\vskip 0.2cm

\vskip 1.7cm
\centerline{{\bf Troels Harmark}\footnote{e-mail: harmark@nbi.dk}}
\vskip 0.3cm
\centerline{\sl The Niels Bohr Institute}
\centerline{\sl Blegdamsvej 17, DK-2100 Copenhagen \O, Denmark}
\vskip 1.7cm
\centerline{\bf Abstract}
\vskip 0.4cm

\noindent

We conjecture the existence of two new non-gravitational
six-dimensional string theories, defined as the decoupling limit
of NS5-branes in the background of near-critical electrical 
two- and three-form RR fields.
These theories are space-time non-commutative Little String Theories 
with open branes.
The theory with $(2,0)$ supersymmetry has an open membrane
in the spectrum and reduces to OM theory at low energies.
The theory with $(1,1)$ supersymmetry has an open string 
in the spectrum and reduces to 5+1 dimensional NCOS theory for 
weak NCOS coupling and low energies.
The theories are shown to be T-dual with
the open membrane being T-dual to the open string.
The theories therefore provide a connection 
between 5+1 dimensional NCOS theory and OM theory.
We study the supergravity duals of these theories and
we consider a chain of dualities that shows how the T-duality
between the two theories is connected with the S-duality between
4+1 dimensional NCOS theory and OM theory.

\end{titlepage}


\newcommand{\nn}{\nonumber}
\newcommand{\spa}{\ \ ,\ \ \ \ }
\newcommand{\str}{\mathop{{\rm Str}}}
\newcommand{\tr}{\mathop{{\rm Tr}}}
\newcommand{\sn}{\mathop{{\rm sn}}}

\newcommand{\gym}{g_{\mathrm{YM}}}
\newcommand{\geff}{g_{\mathrm{eff}}}
\newcommand{\gseff}{g_s^{\mathrm{eff}}}
\newcommand{\Ord}{{\cal{O}}}
\newcommand{\tlst}{T_{\rm LST}}
\newcommand{\tncos}{T_{\rm NCOS}}
\newcommand{\lsb}{\bar{l}_s}
\newcommand{\gsba}{\bar{g}_a}
\newcommand{\gsbb}{\bar{g}_b}
\newcommand{\lsn}{l_s}
\newcommand{\gsna}{g_a}
\newcommand{\gsnb}{g_b}
\newcommand{\lnc}{l_{\rm nc}}
\newcommand{\lm}{l_m}
\newcommand{\uncos}{u_{\rm NCOS}}

\newcommand{\ym}{${\rm YM}_{4+1}$}
\newcommand{\ymm}{${\rm YM}_{5+1}$}
\newcommand{\ncos}{${\rm NCOS}_{4+1}$}
\newcommand{\ncoss}{${\rm NCOS}_{5+1}$}


\setcounter{page}{1}

\tableofcontents

\section{Introduction}

Recently, it has been discovered that the world-volume theory of
a D$p$-brane with a near-critical electrical NSNS $B$-field is a space-time
non-commutative open string (NCOS) theory 
\cite{Seiberg:2000ms,Gopakumar:2000na}. 
Subsequently, it was shown that the world-volume theory of the
M5-brane with a near-critical electrical three-form $C$-field is
a non-commutative open membrane (OM) theory 
\cite{Gopakumar:2000ep,Bergshoeff:2000ai}%
\footnote{For related papers about NCOS theory, OM theory and
space-time non-commutativity, see 
\cite{Seiberg:2000gc}-\cite{Cai:2000yk}.}.
OM theory has been shown 
\cite{Gopakumar:2000ep,Bergshoeff:2000ai,Kawano:2000gn} 
to encompasses all the $p+1$ dimensional
NCOS theories with $p \leq 4$, along with their strong coupling
duals. In this sense, we can see OM theory as a unified framework
for all these lower dimensional theories, in much the same way
as M-theory can be seen as a unified framework of lower dimensional
string theories.
Another close analogy to OM theory is the way that the 5+1 dimensional
$(2,0)$ SCFT encompasses all of the $p+1$ dimensional Yang-Mills (YM) 
theories with $p \leq 4$.

However, the 5+1 dimensional NCOS theory does not appear to be directly
related to OM theory. 
If we use the analogy to $(2,0)$ SCFT and YM theories, we know
that the ultraviolet completion of the 5+1 dimensional YM theory 
is the 5+1 dimensional $(1,1)$ Little String Theory (LST) 
\cite{Berkooz:1997cq,Seiberg:1997zk}%
\footnote{See also \cite{Dijkgraaf:1997hk,Dijkgraaf:1997nb,Losev:1997hx} and 
see \cite{Aharony:1999ks} for a brief review of LST.}
living on the world-volume of type IIB NS5-branes. 
The T-dual of the $(1,1)$ LST is the $(2,0)$ LST living on type IIA 
NS5-branes, and the low energy limit of this theory is the 
$(2,0)$ SCFT. Thus, the two 5+1 dimensional LSTs provide a relation
between 5+1 dimensional YM and $(2,0)$ SCFT.
This also means that we can consider the LSTs as encompassing
both the $(2,0)$ SCFT and the YM theories.

In this paper, we find a relation between 5+1 dimensional NCOS
and OM theory by defining two new theories which we call
$(1,1)$ and $(2,0)$ Open Brane Little String Theories (OBLSTs).
The $(1,1)$ OBLST is defined as the world-volume theory
of $N$ type IIB NS5-branes with a near-critical two-form RR-field,
and the $(2,0)$ OBLST is defined as the world-volume theory
of $N$ type IIA NS5-branes with a near-critical three-form RR-field.
The $(1,1)$ OBLST inherits the closed string from $(1,1)$ LST
but has in addition the open string of 5+1 dimensional NCOS
along with the space-time non-commutativity,
since the decoupling limit of $(1,1)$ OBLST is in fact identical
to that of 5+1 dimensional NCOS, as can be seen from type IIB S-duality.
The $(2,0)$ OBLST has also the closed string of $(2,0)$ LST and
in addition the open membrane of OM theory, again with a
non-commutative geometry. 
Thus, the OBLSTs have open branes and are space-time non-commutative.
For low energies the $(2,0)$ OBLST reduces to OM theory,
while the $(1,1)$ OBLST reduces to 5+1 dimensional NCOS theory
for weak NCOS coupling and low energies.
We show that the $(1,1)$ and $(2,0)$ OBLST are related by T-duality,
in the sense that a T-duality in one of the open membrane directions
in $(2,0)$ OBLST gives the open string of $(1,1)$ OBLST.
The $(1,1)$ and $(2,0)$ OBLST therefore provide a
relation between 5+1 dimensional NCOS and OM theory, and
we can consider them as encompassing OM theory and all the NCOS theories.

In order to explore the OBLSTs we find their supergravity duals.
As part of this we also find the supergravity dual of OM theory.
We subsequently examine the phase structure and thermodynamics of 
the supergravity duals.
From this, we see that the $(1,1)$ OBLST only has an NCOS phase
when the NCOS coupling is small. 
For strong coupling, the closed string from LST dominates.
The $(2,0)$ OBLST reduces to OM theory at low energies
in the supergravity description, as it should.
At high energies the closed string inherited from
LST dominates in both of the OBLSTs, just as for ordinary LST.

We test the consistency of our decoupling/near-horizon 
limits of the OBLSTs by connecting five different bound-states
and their decoupling/near-horizon limits through S- and T-dualities.
The chain of theories we relate is: OM-theory from M2-M5, $D=4+1$ NCOS from
F1-D4, $D=5+1$ NCOS/$(1,1)$ OBLST from F1-D5, $D=5+1$ NCOS/$(1,1)$ OBLST
from D1-NS5 and $(2,0)$ OBLST from D2-NS5. Since the $(2,0)$ OBLST
from D2-NS5 is related directly to OM theory, we have a closed chain
of bound states and limits.
Thus, we can start at any point in the chain and then move on to
other points. 
The S- and T-dualities are also seen to induce corresponding dualities
in the world-volume theories.

It is important to note that instead 
of working in terms of decoupling limits we work mostly with
near-horizon limits in this paper. 
The decoupling limits can easily be read off from the 
near-horizon limits.
Therefore, when considering a particular near-horizon limit
we also consider this limit as defining the theory 
which the corresponding near-horizon supergravity solution
is dual to.

\section{$(1,1)$ OBLST and $D=5+1$ NCOS theory}
\label{secOBLST11}

\subsection{Introduction to $(1,1)$ OBLST}

In \cite{Gopakumar:2000na} a new theory was found from the F1-D5 bound-state 
in the decoupling limit
\begin{equation}
\gsbb \rightarrow \infty \spa
\gsbb \lsb^2 = \mbox{fixed} \spa
B_{01} \rightarrow B_{01}^{\rm critical}
\end{equation}
where $\gsbb$ is the string coupling, $\lsb$ the string length and
$B$ is the two-form NSNS field.
This theory was subsequently shown to be a 
5+1 dimensional theory of open strings, known as 5+1 dimensional NCOS theory, 
living in a space-time geometry with space and time being non-commutative. 

In the following, we shall see that this theory also can be seen as a
space-time non-commutative version of the $(1,1)$ LST.
In fact, using type IIB S-duality we can define the same theory from 
the D1-NS5 bound-state in the decoupling limit 
\begin{equation}
\gsnb \rightarrow 0 \spa 
\lsn = \mbox{fixed} \spa
A_{01} \rightarrow A_{01}^{\rm critical}
\end{equation}
where \( \gsnb = 1 / \gsbb \), \( \lsn^2 = \gsbb \lsb^2 \) 
and $A$ is the RR two-form field.
Thus, just like for ordinary $(1,1)$ LST, the low energy gauge theory
on D1-NS5, which has gauge coupling $\gym^2 = (2\pi)^3 \lsn^2$, should have 
a solitonic string of tension $(2\pi)^2 / \gym^2 = 1 / (2\pi \lsn^2) $. 
Since \( \gsnb = 0 \) in the decoupling limit the string cannot leave the
brane.
In order to study the behavior 
of this LST-string, as we will call it in this paper,
at higher energies, we turn to the supergravity dual description of the
theory and in particular the thermodynamics computed from this. 

As we will explain in the following, 
for weak NCOS coupling and low energies the 
$(1,1)$ OBLST reduces to what we call
$D=5+1$ NCOS, being a theory of weakly coupled open strings.
Thus, $D=5+1$ NCOS can be regarded as a low energy limit of $(1,1)$ OBLST%
\footnote{This was also discussed in \cite{Harmark:2000wv}.}.
On the other hand, when the NCOS coupling is large, the LST tension
is small, so we instead have a space-time non-commutative 
LST governing the dynamics of the theory.
$(1,1)$ OBLST reduces to Yang-Mills theory when the effective Yang-Mills
coupling is small.

\subsection{The F1-D5 and D1-NS5 bound states}

In this section we give the F1-D5 and D1-NS5 bound-states 
so that we can find the supergravity dual description 
of $(1,1)$ OBLST in the next section.

We introduce here the notation that 
the S-dual string couplings and string lengths are connected
as \( \gsnb = 1 / \gsbb \) and \( \lsn^2 = \gsbb \lsb^2 \).
Our notation for the string couplings are further clarified
in section \ref{secchain}.

The non-extremal F1-D5 bound-state has the string frame 
metric \cite{Lu:1999uc,Harmark:2000wv}
\begin{eqnarray}
\label{F1D5met}
ds^2 &=& H^{-1/2} \Big[ D \Big( - f dt^2 + (dx^1)^2 \Big)
+  (dx^2)^2 + \cdots + (dx^5)^2 \Big]
\nn \\ && + H^{1/2} \Big[ f^{-1} dr^2 + r^2 d\Omega_3^2 \Big]
\end{eqnarray}
the dilaton
\begin{equation}
\label{F1D5dil}
e^{2\phi} = H^{-1} D
\end{equation}
and potentials
\begin{eqnarray}
\label{F1D5pot}
B_{01} &=& \sin \hat{\theta} \coth \hat{\alpha} D H^{-1} 
\\
A_{2345} &=& - \tan \hat{\theta} H^{-1}
\\
A_{012345} &=& - \frac{1}{\cos \hat{\theta}} \coth \hat{\alpha} (H^{-1}-1)
\end{eqnarray}
with $B_{\mu \nu}$ being the NSNS two-form field, 
$A_{\mu \nu \rho \sigma}$ being the RR four-form field
and $A_{\mu \nu \rho \sigma \kappa \lambda}$ being the RR six-form field.
We also define
\begin{eqnarray}
\label{F1D5defs}
f &=& 1 - \frac{r_0^2}{r^2}
\\
H &=& 1 + \frac{r_0^2 \sinh^2 \alpha}{r^2} 
\\
D^{-1} &=& \cosh^2 \theta - \sinh^2 \theta H^{-1}
\end{eqnarray}
We use the two sets of variables $\theta$, $\alpha$ and $\hat{\theta}$,
$\hat{\alpha}$ related by
\begin{equation}
\label{newvar}
\sinh^2 \alpha = \cos^2 \hat{\theta} \sinh^2 \hat{\alpha} \spa
\cosh^2 \theta = \frac{1}{\cos^2 \hat{\theta}}
\end{equation}
Using charge quantization of the $N$ D5-branes we get
\begin{equation}
r_0^2 \sinh{\alpha} \sqrt{ \sinh^2 \alpha + \cosh^{-2} \theta }
= \frac{\gsbb \lsb^2 N}{\cosh \theta}
= \frac{\lsn^2 N}{\cosh \theta}
\end{equation}

We now use type IIB S-duality on the F1-D5 solution 
\eqref{F1D5met}-\eqref{F1D5pot}.
This gives the D1-NS5 solution
\begin{eqnarray}
\label{D1NS5met}
ds^2 &=& D^{-1/2} \left[ D \Big( - f dt^2 + (dx^1)^2 \Big)
+ (dx^2)^2 + \cdots + (dx^5)^2 
\right. \nn \\ && \left.
+ H \Big( f^{-1} dr^2 + r^2 d\Omega_3^2 \Big) \right]
\end{eqnarray}
\begin{equation}
\label{D1NS5dil}
e^{2\phi} = H D^{-1}
\end{equation}
\begin{eqnarray}
\label{D1NS5pot}
A_{01} &=& - \sin \hat{\theta} \coth \hat{\alpha} D H^{-1} 
\\
A_{2345} &=& - \tan \hat{\theta} H^{-1}
\end{eqnarray}

\subsection{Supergravity description of $(1,1)$ OBLST}
\label{secSupOB11}

The near-horizon limit of the F1-D5 bound-state is 
\cite{Gopakumar:2000na,Harmark:2000wv}
\begin{equation}
\lsb \rightarrow 0 \spa
\gsbb \lsb^2 = \mbox{fixed} \spa
\tilde{r} = \frac{\sqrt{b}}{\lsb} r \spa
\tilde{r}_0 = \frac{\sqrt{b}}{\lsb} r_0 \spa
b = \lsb^2 \cosh \theta
\end{equation}
\begin{equation}
\tilde{x}^i = \frac{\lsb}{\sqrt{b}} x^i \ , \ i=0,1 \spa
\tilde{x}^j = \frac{\sqrt{b}}{\lsb} x^j \ , \ j=2,...,5 
\end{equation}
where we use the notation \( x^0 = t \).
We have
\begin{equation}
L^2 = \tilde{r}_0^2 \sinh^2 \alpha = \gsbb \lsb^2 N = \lsn^2 N
\end{equation}
We get the near-horizon solution \cite{Harmark:2000wv}
\begin{eqnarray}
\label{NHF1D5met}
ds^2 &=& \frac{\lsb^2}{b} H^{-1/2} \left[ H \frac{\tilde{r}^2}{L^2} 
\Big( - f d\tilde{t}^2 + (d\tilde{x}^1)^2 \Big)
+ (d\tilde{x}^2)^2 + \cdots + (d\tilde{x}^5)^2 
\right. \nn \\ && \left.
+ H \Big( f^{-1} d\tilde{r}^2 + \tilde{r}^2 d\Omega_3^2 \Big) \right]
\end{eqnarray}
\begin{equation}
\label{NHF1D5dil}
\gsbb^2 e^{2\phi} 
= \frac{\lsn^4}{b^2} \frac{\tilde{r}^2}{L^2}
\end{equation}
\begin{equation}
\label{NHF1D5pot}
B_{01} = \frac{\lsb^2}{b} \frac{\tilde{r}^2}{L^2}
\end{equation}
with
\begin{equation}
\label{NHF1D5defs}
f = 1 - \frac{\tilde{r}_0^2}{\tilde{r}^2} \spa
H = 1 + \frac{L^2}{\tilde{r}^2}
\end{equation}

Using S-duality, the near-horizon limit of D1-NS5 is
\begin{equation}
\label{D1NS5lim1}
\gsnb \rightarrow 0 \spa
\lsn = \mbox{fixed} \spa
\tilde{r} = \frac{\sqrt{b}}{\sqrt{\gsnb} \lsn} r \spa
\tilde{r}_0 = \frac{\sqrt{b}}{\sqrt{\gsnb} \lsn} r_0 \spa
b = \gsnb \lsn^2 \cosh \theta 
\end{equation}
\begin{equation}
\label{D1NS5lim2}
\tilde{x}^i = \frac{\sqrt{\gsnb} \lsn}{\sqrt{b}} x^i \ , \ i=0,1 \spa
\tilde{x}^j = \frac{\sqrt{b}}{\sqrt{\gsnb} \lsn} x^j \ , \ j=2,...,5 
\end{equation}
The limit \eqref{D1NS5lim1}-\eqref{D1NS5lim2} 
gives the near-horizon solution%
\footnote{The string metric does not go to zero in our notation since
we define the string metric via \( e^{\phi} \) instead 
of \( \gsnb e^{\phi} \).}
\begin{eqnarray}
\label{NHD1NS5met}
ds^2 &=& H^{-1/2} \frac{L}{\tilde{r}} \left[ H \frac{\tilde{r}^2}{L^2} 
\Big( - f d\tilde{t}^2 + (d\tilde{x}^1)^2 \Big)
+ (d\tilde{x}^2)^2 + \cdots + (d\tilde{x}^5)^2 
\right. \nn \\ && \left.
+ H \Big( f^{-1} d\tilde{r}^2 + \tilde{r}^2 d\Omega_3^2 \Big) \right]
\end{eqnarray}
\begin{equation}
\label{NHD1NS5dil}
\gsnb^2 e^{2\phi} 
= \frac{b^2}{\lsn^4} \frac{L^2}{\tilde{r}^2}
\end{equation}
\begin{equation}
\label{NHD1NS5pot}
A_{01} = - \frac{\gsnb \lsn^2}{b} \frac{\tilde{r}^2}{L^2}
\end{equation}

We now give the mapping from our supergravity parameters to the
parameters of $(1,1)$ OBLST.
The $(1,1)$ OBLST lives on a non-commutative space-time with the commutator
$[\tilde{t},\tilde{x}^1]=ib$.
The energy coordinate $u$ is related to the rescaled radial
coordinate $\tilde{r}$ as $u = \tilde{r}/b$.
As we discuss in the following, the $(1,1)$ OBLST has three different
phases: A weakly coupled Yang-Mills phase, 
a weakly coupled NCOS phase and a phase with LST strings. 
There are special parameters for each of these phases.

The open string coupling of the 5+1 dimensional NCOS 
is \cite{Gopakumar:2000na}%
\footnote{For convinience we call $\tilde{g} = G_o^2$ 
the NCOS open string coupling where $G_o$ is the NCOS open string coupling
of \cite{Gopakumar:2000na}.}
\begin{equation}
\tilde{g} = \frac{\gsbb \lsb^2}{b} = \frac{\lsn^2}{b}
\end{equation}
The tension of the open string is $1/b$.
The Yang-Mills coupling constant of the 5+1 dimensional Yang-Mills theory is
$\gym^2 = (2\pi)^3 \tilde{g} b $. This gives the
effective YM coupling constant $\geff^2 = (2\pi)^3 \tilde{g} N b u^2 $.
The LST-string has the tension \( 1/(2\pi \lsn^2 ) \).

\begin{figure}[h]
\begin{picture}(390,75)(0,0)
\put(5,40){\vector(1,0){380}}
\put(382,50){$u$}
\put(20,5){\shortstack{\ymm \\ Perturbative}}
\put(115,5){\shortstack{$(1,1)$ LST \\ D5-branes}}
\put(210,5){\shortstack{$(1,1)$ LST \\ NS5-branes}}
\put(300,5){\shortstack{$(1,1)$ OBLST \\ D-strings}}
\put(2,50){0}
\put(5,35){\line(0,1){10}}
\put(85,55){$\frac{1}{\lsn \sqrt{N}}$}
\put(96,35){\line(0,1){10}}
\put(182,55){$\frac{\sqrt{N}}{\lsn}$}
\put(192,35){\line(0,1){10}}
\put(276,55){$\frac{\sqrt{N}\lsn}{b}$}
\put(288,35){\line(0,1){10}}
\end{picture}
\caption{Phase diagram for $(1,1)$ OBLST with 
\( \tilde{g} \gg 1 \). \label{figOB11c} }
\end{figure}
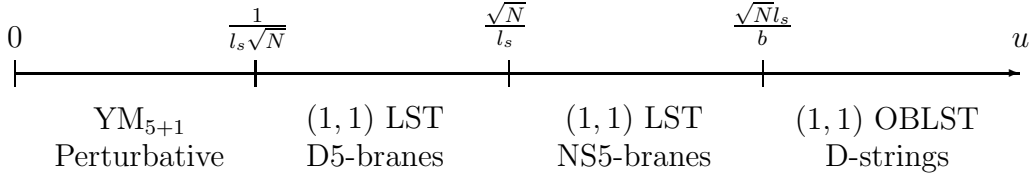

\begin{figure}[h]
\begin{picture}(390,75)(0,0)
\put(5,40){\vector(1,0){380}}
\put(382,50){$u$}
\put(35,5){\shortstack{\ncoss \\ Perturbative}}
\put(155,5){\shortstack{$(1,1)$ OBLST \\ F-strings}}
\put(280,5){\shortstack{$(1,1)$ OBLST \\ D-strings}}
\put(2,50){0}
\put(5,35){\line(0,1){10}}
\put(120,55){$\frac{1}{\sqrt{b}}$}
\put(127,35){\line(0,1){10}}
\put(244,55){$\frac{\sqrt{N}}{\lsn}$}
\put(254,35){\line(0,1){10}}
\end{picture}
\caption{Phase diagram for $(1,1)$ OBLST with 
\( \tilde{g} \ll 1/N \). \label{figOB11a} }
\end{figure}
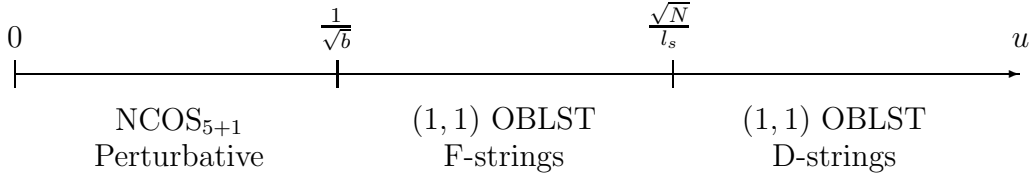

We now consider the phase structure of $(1,1)$ OBLST in terms
of phase diagrams with the energy coordinate $u$ as variable.
The three possible phase diagrams for $N \gg 1$ are depicted in
figure \ref{figOB11b}-\ref{figOB11a}. For $N \sim 1$ we have instead only two
phase diagrams.

\begin{figure}[h]
\begin{picture}(390,75)(0,0)
\put(5,40){\vector(1,0){380}}
\put(382,50){$u$}
\put(20,5){\shortstack{\ymm \\ Perturbative}}
\put(115,5){\shortstack{$(1,1)$ LST \\ D5-branes}}
\put(205,5){\shortstack{$(1,1)$ OBLST \\ F-strings}}
\put(300,5){\shortstack{$(1,1)$ OBLST \\ D-strings}}
\put(2,50){0}
\put(5,35){\line(0,1){10}}
\put(85,55){$\frac{1}{\lsn \sqrt{N}}$}
\put(96,35){\line(0,1){10}}
\put(180,55){$\frac{\sqrt{N}\lsn}{b}$}
\put(192,35){\line(0,1){10}}
\put(278,55){$\frac{\sqrt{N}}{\lsn}$}
\put(288,35){\line(0,1){10}}
\end{picture}
\caption{Phase diagram for $(1,1)$ OBLST with 
\( 1/N \ll \tilde{g} \ll 1 \). \label{figOB11b} }
\end{figure}
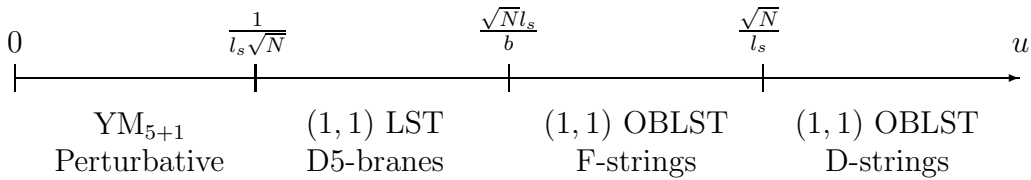

We observe that the supergravity dual of $(1,1)$ OBLST
reduces to the supergravity dual of $(1,1)$ LST given in 
\cite{Itzhaki:1998dd,Aharony:1998ub} when $u \ll \sqrt{N} \lsn / b$. 

We consider first \( \tilde{g} N \gg 1 \) which gives the two possible
phase diagrams depicted in figure \ref{figOB11b} and \ref{figOB11c}.
We have three transition points.
At $\geff^2 \sim 1$, which is equivalent to $u \sim 1 / (\lsn \sqrt{N}) $, 
we flow from a perturbative YM description to a near-horizon
D5-brane description.
At $\gsnb e^{\phi} \sim 1$, which is equivalent
to $u \sim \sqrt{N}/\lsn $, we go either from a D5 to a NS5 description,
or from a delocalized F-string to a delocalized D-string description.
At $u \sim L/b = \sqrt{N} \lsn / b $ 
we flow from the ordinary $(1,1)$ LST to $(1,1)$ OBLST, 
and we go either from a D5 to a delocalized F-string description,
or from a NS5 to a delocalized D-string description. 

The third possible phase diagram, depicted in figure \ref{figOB11a},
has \( \tilde{g} N \ll 1 \).
At energies \( u \ll 1 / \sqrt{b} \) we have 
a weakly coupled 5+1 dimensional NCOS theory description,
which reduces to perturbative YM theory at low energies.
At \( u \sim 1/\sqrt{b} \) we flow to a delocalized F-string description
and at \( u \sim \sqrt{N} / \lsn \) we flow to a delocalized D-string
description.

In order to understand these phase diagrams, it is useful first to
consider the thermodynamics of the supergravity 
description.
The near-horizon solutions \eqref{NHF1D5met}-\eqref{NHF1D5pot} 
and \eqref{NHD1NS5met}-\eqref{NHD1NS5pot}
give the leading order thermodynamics \cite{Harmark:2000wv}
\begin{equation}
\label{OB11therm1}
T = \frac{1}{2\pi \lsn \sqrt{N}}  \spa
S = \sqrt{N} \frac{\tilde{V}_5 }{(2\pi)^4} \frac{1}{b^2 \lsn^3} \tilde{r}_0^2
\end{equation}
\begin{equation}
\label{OB11therm2}
E= \frac{\tilde{V}_5 }{(2\pi)^5} \frac{1}{b^2 \lsn^4} \tilde{r}_0^2
\spa 
F= 0
\end{equation}
This thermodynamics describes $(1,1)$ OBLST for 
\( u \gg u_{\rm SG} \),
where $u_{\rm SG} =  1/(\lsn \sqrt{N})$ for $\tilde{g}N \gg 1$ and
$u_{\rm SG} = 1/\sqrt{b}$ for $\tilde{g}N \ll 1$, 
since the string corrections to the
thermodynamics are small in this region.
The Hagedorn temperature of ordinary $(1,1)$ LST is
\begin{equation}
\tlst = \frac{1}{2\pi \lsn \sqrt{N}}
\end{equation}
So, we have that \( T \sim \tlst \) for \( u \gg u_{\rm SG} \).
This suggest that the LST-strings dominates the dynamics of $(1,1)$ OBLST
for $ u \gg u_{\rm SG} $, since the thermodynamics
\eqref{OB11therm1}-\eqref{OB11therm2} has the same 
leading order Hagedorn behaviour as ordinary $(1,1)$ LST 
\cite{Maldacena:1996ya,Maldacena:1997cg,Harmark:2000hw}.
That the LST-strings live on a space-time non-commutative geometry
can be seen by the fact that the critical behavior of the
entropy at very high energies are different, as shown 
in \cite{Harmark:2000wv}.
For ordinary $(1,1)$ LST we have that \cite{Harmark:2000hw,Berkooz:2000mz} 
$S(T) \propto (\tlst -T )^{-1}$
while for $(1,1)$ OBLST we have \cite{Harmark:2000wv} 
$S(T) \propto (\tlst - T)^{-2/3}$.

That the LST-string dominates for $u \gg u_{\rm SG}$ 
is not in contradiction with the existence of open strings
in $(1,1)$ OBLST, as we now explain.

Consider first the case \( 1/N \ll \tilde{g} \ll 1 \) with phase
diagram depicted in figure \ref{figOB11b}.
This case corresponds to strongly coupled open strings, since 
\( \tilde{g} N \gg 1 \).
Though the LST-strings are not lighter in this case, there
is the other energy scale $1/(\lsn \sqrt{N})$ in LST
which is connected with LST Hagedorn behavior. 
As suggested in \cite{Maldacena:1996ya,Harmark:2000hw}, 
this could be a LST-string scale
connected with fractional strings. 
The LST-modes corresponding to this scale are clearly lighter than
the open string modes, thus explaining why the
LST-string modes dominates for $u \gg 1/(\lsn \sqrt{N})$.

The second case has \( \tilde{g} \gg 1 \) 
which also corresponds to strongly coupled open strings.
The phase diagram is depicted in figure \ref{figOB11c}.
We have that $1/b \gg 1/\lsn^2$ thus the LST-strings are lighter
than the open strings and are therefore expected to dominate,
which is confirmed the thermodynamics.

The final case has $ \tilde{g} N \ll 1 $
corresponding to the phase diagram in figure \ref{figOB11a}.
Since we have that $1/b \ll 1/(N\lsn^2) $ the open strings of NCOS theory
are much lighter than the LST-modes connected with LST Hagedorn behaviour 
and we should therefore expect them to dominate the dynamics.
For the NCOS Hagedorn temperature $\tncos$ we know that 
\( \tncos \sim 1/\sqrt{b} \).
Thus, we have that $\tncos \ll \tlst$.
Clearly, the NCOS Hagedorn temperature is not limiting,
and we should have a Hagedorn phase transition at a certain energy
$\uncos$.
Since we have just shown that for $u \gg 1/\sqrt{b}$
we had $T \sim \tlst$, we get that $\uncos \lesssim 1/\sqrt{b}$.
Thus, the reason that the LST-strings can dominate at high energies, 
even though they are heavier than the open strings, is that 
the open strings have been subject to a Hagedorn transition
at these energies.
We note that the lower dimensional NCOS theories exhibit similar behaviour
in the sense that, when they are weakly coupled, the supergravity dual
describes them only when the temperature are above the NCOS Hagedorn
temperature and a Hagedorn phase transition has occured \cite{Harmark:2000wv}.

In summary, we have learned that at weak NCOS coupling $\tilde{g} N \ll 1$ 
the $(1,1)$ OBLST has an NCOS phase for energies
$u \ll 1/\sqrt{b} $, with LST-strings dominating at higher energies, 
as depicted in figure \ref{figOB11a}.
For strong NCOS coupling $\tilde{g} N \gg 1$ we have
perturbative YM for low energies and LST-strings at high energies
as depicted in figure \ref{figOB11b} and \ref{figOB11c}.
Thus, in this sense one can say that strongly coupled 5+1 dimensional
NCOS theory gives a space-time non-commutative version of $(1,1)$ LST.

\subsection{Branes in $(1,1)$ OBLST}

In the ordinary $(1,1)$ LST we have besides the LST-string with
tension $1/(2\pi \lsn^2)$ the d0, d2 and d4 branes \cite{Losev:1997hx}.
These origins from having open D1, D3 and D5 branes
stretching between NS5-branes.

In the $(1,1)$ OBLST we still have the LST-string, but now the
D-string stretching between two D1-NS5 bound-states induces
an open string. 
We note that the zero modes of the open D-string
is what gives the Yang-Mills
theory at low energies, which fits with the picture that the 
NCOS theory at low energies reduce to Yang-Mills theory.

Since there is not any electric field on the ends of an open D3-brane
stretching between D1-NS5 bound-states
 we expect that we still have the same d-membrane
in $(1,1)$ OBLST as in $(1,1)$ LST, but presumably now moving in a 
space-time non-commutative geometry.
Also the d4-brane seem to be part of $(1,1)$ OBLST.

In other words, only the open D-brane for which the potential goes to
its critical value, gives an open brane in the world-volume theory.
The rest of the spectrum is unchanged.

\section{Supergravity dual of OM theory}
\label{secOMdual}

In this section we find and study the supergravity dual of
OM theory.
We find the OM supergravity dual by uplifting the 4+1 dimensional
NCOS supergravity dual.
Apart from being interesting in its own right, it is also
important to understand the OM theory near-horizon limit 
in order to understand the decoupling and near-horizon 
limit of $(2,0)$ OBLST. 

The near-horizon and decoupling limits of the M2-M5 bound state
have previously been studied in \cite{Seiberg:1999vs,Maldacena:1999mh,Alishahiha:1999ci,Harmark:1999rb,Chakravarty:2000qd,Bergshoeff:2000jn,Kawamoto:2000zt,Gopakumar:2000ep,Bergshoeff:2000ai}.

\subsection{The M2-M5 and F1-D4 bound states}
\label{secM2M5F1D4}

In this section we describe the supergravity solutions that 
we use for OM theory and 4+1 dimensional NCOS theory.

The non-extremal M2-M5 brane bound-state has the metric \cite{Russo:1997if}
\begin{eqnarray}
ds_{11}^2 &=& (\hat{H}\hat{D})^{-1/3} \Big[ - f dt^2 +(dx^1)^2 + (dx^2)^2
+ \hat{D} \Big( (dx^3)^2 +(dx^4)^2 + (dx^5)^2 \Big)
\nn \\ &&
+ \hat{H} \Big( f^{-1} dr^2 + r^2 d\Omega_4^2 \Big) \Big]
\end{eqnarray}
and the three- and six-form potentials
\begin{eqnarray}
\label{M2M5pot1}
C_{012} &=& - \sin \hat{\theta} \hat{H}^{-1} \coth \hat{\alpha}
\\
C_{345} &=& \tan \hat{\theta} \hat{D} \hat{H}^{-1}
\\
\label{M2M5pot3}
C_{012345} &=& \cos \hat{\theta} \hat{D} (\hat{H}^{-1} - 1 ) \coth \hat{\alpha}
\end{eqnarray}
We have
\begin{equation}
f = 1 - \frac{r_0^3}{r^3} \spa
\hat{H} = 1 + \frac{r_0^3 \sinh^2 \hat{\alpha}}{r^3}
\end{equation}
\begin{equation}
\hat{D}^{-1} = \cos^2 \hat{\theta} + \sin^2 \hat{\theta} \hat{H}^{-1}
\end{equation}
The charge quantization for $N$ M5-branes gives
\begin{equation}
r_0^3 \cosh \hat{\alpha} \sinh \hat{\alpha} 
= \frac{\pi N l_p^3}{\cos \hat{\theta}}
\end{equation}
In the new variables 
\begin{equation}
\sinh^2 \alpha = \cos^2 \hat{\theta} \sinh^2 \hat{\alpha} \spa
\cosh^2 \theta = \frac{1}{\cos^2 \hat{\theta}}
\end{equation}
we have
\begin{eqnarray}
\label{M2M5met}
ds_{11}^2 &=& (HD)^{-1/3} \Big[ D \Big( - f dt^2 +(dx^1)^2 + (dx^2)^2 \Big)
+ (dx^3)^2 +(dx^4)^2 + (dx^5)^2 
\nn \\ &&
+ H \Big( f^{-1} dr^2 + r^2 d\Omega_4^2 \Big) \Big]
\end{eqnarray}
\begin{eqnarray}
H &=& 1 + \frac{r_0^3 \sinh^2 \alpha}{r^3}
\\
D^{-1} &=& \cosh^2 \theta - \sinh^2 \theta H^{-1}
\end{eqnarray}
We note that \( \hat{H} = H D^{-1} \), \( H = \hat{H} \hat{D}^{-1} \)
and \( D = \hat{D}^{-1} \).

We now dimensionally reduce the M2-M5 on the electric circle
with the coordinate \( x^2 \).
This gives the F1-D4 bound-state solution.
The relation between the eleven dimensional metric
$ds_{11}^2$ and the ten-dimensional string-frame metric $ds_{10}^2$
and dilaton $e^{\phi}$ is
\begin{equation}
ds_{11}^2 = e^{-\frac{2}{3} \phi} ds_{10}^2 + e^{\frac{4}{3}\phi} (dx^2)^2
\end{equation}
Thus, we get the metric
\begin{eqnarray}
\label{F1D4met}
ds_{10}^2 &=& H^{-1/2} \Big[ D \Big( - f dt^2 + (dx^1)^2 \Big)
+  (dx^3)^2 + (dx^4)^2 + (dx^5)^2 
\nn \\ && + H \Big( f^{-1} dr^2 + r^2 d\Omega_4^2 \Big) \Big]
\end{eqnarray}
the dilaton
\begin{equation}
\label{F1D4dil}
e^{2\phi} = H^{-1/2} D
\end{equation}
and the NSNS two-form potential
\begin{equation}
\label{F1D4pot}
B_{01} = - \sin \hat{\theta} \coth \hat{\alpha} D H^{-1}
\end{equation}
This solution coincides with the one given in \cite{Harmark:2000wv}.

\subsection{Near-horizon limit of OM theory from $D=4+1$ NCOS}
\label{secOMlimit}

The near horizon limit of 4+1 dimensional NCOS 
is \cite{Gopakumar:2000na,Harmark:2000wv}
\begin{equation}
\label{F1D4lim1}
\lsb \rightarrow 0 \spa
\tilde{r} = \frac{\sqrt{b}}{\lsb} r \spa
\tilde{r}_0 = \frac{\sqrt{b}}{\lsb} r_0 \spa
b = \lsb^2 \cosh \theta \spa
\alpha = \mbox{fixed}
\end{equation}
\begin{equation}
\label{F1D4lim2}
\tilde{x}^i = \frac{\lsb}{\sqrt{b}} x^i \ , \ i=0,1 \spa
\tilde{x}^j = \frac{\sqrt{b}}{\lsb} x^j \ , \ j=3,4,5 
\end{equation}
We have the open string coupling squared 
\begin{equation}
\tilde{g} = \frac{\gsba \lsb^2}{b}
\end{equation}
Using \eqref{F1D4lim1}-\eqref{F1D4lim2} 
on \eqref{F1D4met}-\eqref{F1D4pot} we get \cite{Harmark:2000wv}
\begin{eqnarray}
ds_{10}^2 &=& \frac{\lsb^2}{b} H^{-1/2} \left[ 
H \frac{\tilde{r}^2}{L^2} \Big( - f d\tilde{t}^2 + (d\tilde{x}^1)^2 \Big) 
+ (d\tilde{x}^3)^2 + (d\tilde{x}^4)^2 + (d\tilde{x}^5)^2
\right. \nn \\ && \left.
+ H \Big( f^{-1} d\tilde{r}^2 + \tilde{r}^2 d\Omega_4^2 \Big) \right]
\end{eqnarray}
\begin{equation}
\gsba^2 e^{2\phi} = \tilde{g}^2 H^{-1/2} 
\Big( 1 + \frac{\tilde{r}^3}{L^3} \Big)
\end{equation}
\begin{equation}
B_{01} = \frac{\lsb^2}{b} \frac{\tilde{r}^3}{L^3}
\end{equation}
with
\begin{equation}
L^3 = \tilde{r}_0^3 \sinh^2 \alpha = \pi N \tilde{g} b^{3/2}
\end{equation}
\begin{equation}
H = 1 + \frac{L^3}{\tilde{r}^3} \spa
f = 1 - \frac{\tilde{r}_0^3}{\tilde{r}^3}
\end{equation}

We now introduce the open membrane length scale \( \lm \) in OM theory
defined by stating that the open membrane has tension \( 1 / \lm^3 \).
As shown in \cite{Gopakumar:2000ep,Bergshoeff:2000ai} we then have
\begin{equation}
\label{lmcube}
\lm^3 = \tilde{g} b^{3/2}
\end{equation}
This can be understod as follows.
Since the radius of the electric circle with coordinate \( x^2 \)
is \( R_E = \gsba \lsb \) the rescaled radius is 
\begin{equation}
\tilde{R}_E = \frac{\gsba \lsb^2}{\sqrt{b}} = \tilde{g} \sqrt{b}
\end{equation}
where the rescaling \( \tilde{R}_E = R_E \lsb / \sqrt{b} \) follows
from the fact that the \( x^2 \) coordinate should scale the same
way as the \( x^1 \) coordinate in \eqref{F1D4lim2}.
We can then write the relation%
\footnote{In these type of relations we ignore factors of $2\pi$.}
\begin{equation}
\frac{1}{b} = \frac{\tilde{R}_E}{\lm^3}
\end{equation}
where \( 1/b \) is the tension of the open string in NCOS.
Thus, the relation \eqref{lmcube} is the statement 
that the open string in NCOS is the open membrane in OM theory
wrapped on a circle of radius \( \tilde{R}_E \).

We now want to use \eqref{lmcube} to write the near-horizon limit 
\eqref{F1D4lim1}-\eqref{F1D4lim2} in terms of the eleven dimensional
variables \( \lm \) and \( l_p \), where \( l_p^3 = \gsba \lsb^3 \).
Using \eqref{lmcube} we have 
\begin{equation}
\frac{\sqrt{b}}{\lsb} = \frac{\tilde{g} b^{3/2}}{\tilde{g} b \lsb}
= \frac{\lm^3}{l_p^3}
\end{equation}
Using this together with \eqref{F1D4lim1}-\eqref{F1D4lim2} 
we can write the eleven dimensional near-horizon limit of
OM theory as
\begin{equation}
\label{M2M5lim1}
l_p \rightarrow 0 \spa
\tilde{r} = \frac{\lm^3}{l_p^3} r \spa
\tilde{r}_0 = \frac{\lm^3}{l_p^3 } r_0 \spa
\lm^6 = l_p^6 \cosh \theta
\end{equation}
\begin{equation}
\label{M2M5lim2}
\tilde{x}^i = \frac{l_p^3}{\lm^3} x^i, \ \ i=0,1,2 \spa
\tilde{x}^j = \frac{\lm^3}{l_p^3} x^j,\ \ j=3,4,5 
\end{equation}
This is a purely eleven dimensional near-horizon limit of
OM theory, meaning that it can describe OM theory
with the Lorentz symmetry \( SO(1,2) \times SO(3) \).

The near-horizon limit \eqref{M2M5lim1}-\eqref{M2M5lim2} 
is the same limit of the M2-M5 brane bound state
as in \cite{Alishahiha:1999ci,Harmark:1999rb}.
Keeping $r/l_p^3$ fixed means that the membrane modes for open
M2-branes stretching between M5-branes are kept finite.

Using \eqref{M2M5lim1}-\eqref{M2M5lim2} on
\eqref{M2M5met} and \eqref{M2M5pot1}-\eqref{M2M5pot3} 
we get the supergravity dual
\begin{eqnarray}
\label{NHM2M5met}
ds_{11}^2 &=& \frac{l_p^2}{\lm^2} H^{-2/3} \frac{L}{\tilde{r}} \left[
H \frac{\tilde{r}^3}{L^3} \Big( - f d\tilde{t}^2 
+ (d\tilde{x}^1)^2 + (d\tilde{x}^2)^2 \Big)
\right. \nn \\ && \left.
+ (d\tilde{x}^3)^2 + (d\tilde{x}^4)^2 + (d\tilde{x}^5)^2 
+ H \Big( f^{-1} d\tilde{r}^2 + \tilde{r}^2 d\Omega_4^2 \Big) \right]
\end{eqnarray}
\begin{equation}
\label{NHM2M5pot}
C_{012} = - \frac{l_p^3}{\lm^3} \frac{\tilde{r}^3}{L^3}
\spa
C_{345} = \frac{l_p^3}{\lm^3} H^{-1}
\end{equation}
\begin{equation}
\label{NHM2M5defs}
H = 1 + \frac{L^3}{\tilde{r}^3} \spa
f = 1 - \frac{\tilde{r}_0^3}{\tilde{r}^3}
\end{equation}

When OM theory is on an electric circle of 
rescaled radius $\tilde{R}_E$, the energy coordinate $u$ is
\begin{equation}
u = \frac{\tilde{r}}{b} = \frac{\tilde{R}_E \tilde{r}}{\lm^3}
\end{equation}

\subsection{Phase structure of OM and $D=4+1$ NCOS theory}
\label{secOMphases}

In this section we examine the phase structure of OM theory
and the 4+1 dimensional NCOS theory via their supergravity duals.

The OM theory near-horizon solution \eqref{NHM2M5met}-\eqref{NHM2M5pot} 
is valid when the curvature in units of \( l_p^{-2} \) 
\begin{equation}
\label{OMcurv}
\mathcal{C} = \left( \pi N^2 + \frac{N \tilde{r}^3}{\lm^3} \right)^{-1/3}
\end{equation}
is small.
Thus, if $N \gg 1$ this is clearly satisfied and we can describe 
OM theory for all energies. 
If $N$ is of order 1, we instead need that \( \tilde{r} \gg N^{1/3} \lm \)
since we need that \( \tilde{r} \gg L \).

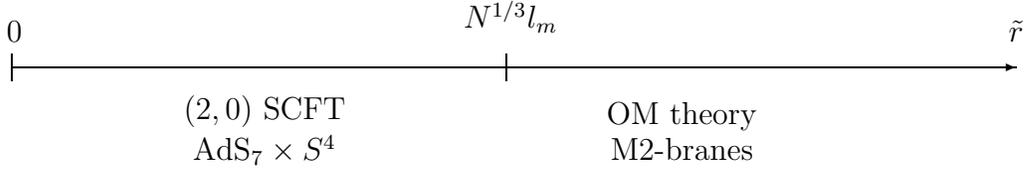
\begin{figure}[h]
\begin{picture}(390,75)(0,0)
\put(5,40){\vector(1,0){380}}
\put(382,50){$\tilde{r}$}
\put(70,5){\shortstack{$(2,0)$ SCFT \\ ${\rm AdS}_7 \times S^4$ }}
\put(230,5){\shortstack{OM theory \\ M2-branes}}
\put(3,50){0}
\put(5,35){\line(0,1){10}}
\put(176,55){$N^{1/3} \lm$}
\put(192,35){\line(0,1){10}}
\end{picture}
\caption{Phase diagram of OM theory. \label{figOMphases} }
\end{figure}

From the supergravity dual \eqref{NHM2M5met}-\eqref{NHM2M5pot} we see 
that for $\tilde{r} \ll N^{1/3} \lm$ the solution reduces to ${\rm AdS}_7 \times S^4$
describing the six-dimensional $(2,0)$ SCFT \cite{Maldacena:1997re}.
Thus, for $\tilde{r} \ll N^{1/3} \lm$ OM theory reduces to $(2,0)$ SCFT.
For $\tilde{r} \gg N^{1/3} \lm$ 
the open membrane is large enough to have an
effect and the underlying non-commutative geometry is detectable.
For $\tilde{r} \gg N^{1/3} \lm$ the solution is described by
M2-branes delocalized in 3 directions.
The phases are depicted in Figure \ref{figOMphases}.

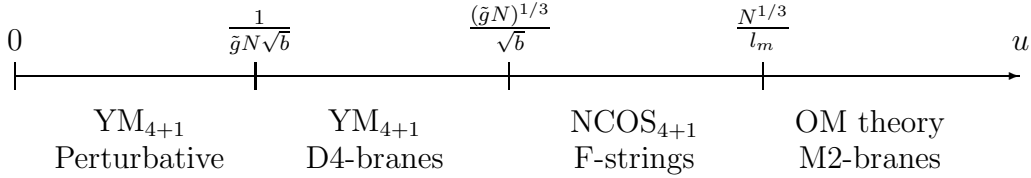
\begin{figure}[h]
\begin{picture}(390,75)(0,0)
\put(5,40){\vector(1,0){380}}
\put(382,50){$u$}
\put(20,5){\shortstack{\ym \\ Perturbative}}
\put(115,5){\shortstack{\ym \\ D4-branes}}
\put(215,5){\shortstack{\ncos \\ F-strings}}
\put(300,5){\shortstack{OM theory \\ M2-branes}}
\put(2,50){0}
\put(5,35){\line(0,1){10}}
\put(85,55){$\frac{1}{\tilde{g} N\sqrt{b}}$}
\put(96,35){\line(0,1){10}}
\put(176,55){$\frac{(\tilde{g} N)^{1/3}}{\sqrt{b}}$}
\put(192,35){\line(0,1){10}}
\put(277,55){$\frac{N^{1/3}}{\lm}$}
\put(288,35){\line(0,1){10}}
\end{picture}
\caption{Phase diagram for $D=4+1$ NCOS with 
\( 1/N \ll \tilde{g} \ll 1 \). \label{figF1D4ph1} }
\end{figure}

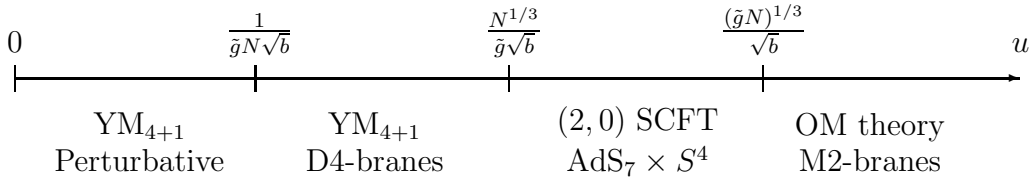
\begin{figure}[h]
\begin{picture}(390,75)(0,0)
\put(5,40){\vector(1,0){380}}
\put(382,50){$u$}
\put(20,5){\shortstack{\ym \\ Perturbative}}
\put(115,5){\shortstack{\ym \\ D4-branes}}
\put(210,5){\shortstack{$(2,0)$ SCFT \\ ${\rm AdS}_7 \times S^4$}}
\put(300,5){\shortstack{OM theory \\ M2-branes}}
\put(2,50){0}
\put(5,35){\line(0,1){10}}
\put(85,55){$\frac{1}{\tilde{g} N\sqrt{b}}$}
\put(96,35){\line(0,1){10}}
\put(183,55){$\frac{N^{1/3}}{\tilde{g} \sqrt{b}}$}
\put(192,35){\line(0,1){10}}
\put(272,55){$\frac{(\tilde{g}N)^{1/3}}{\sqrt{b}}$}
\put(288,35){\line(0,1){10}}
\end{picture}
\caption{Phase diagram for $D=4+1$ NCOS with
\( \tilde{g} \gg 1 \). \label{figF1D4ph2} }
\end{figure}

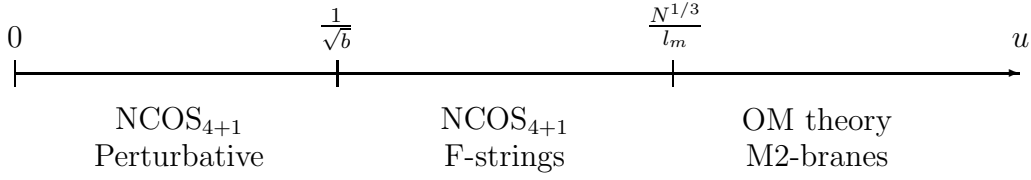
\begin{figure}[h]
\begin{picture}(390,75)(0,0)
\put(5,40){\vector(1,0){380}}
\put(382,50){$u$}
\put(35,5){\shortstack{\ncos \\ Perturbative}}
\put(166,5){\shortstack{\ncos \\ F-strings}}
\put(280,5){\shortstack{OM theory \\ M2-branes}}
\put(2,50){0}
\put(5,35){\line(0,1){10}}
\put(120,55){$\frac{1}{\sqrt{b}}$}
\put(127,35){\line(0,1){10}}
\put(244,55){$\frac{N^{1/3}}{\lm}$}
\put(254,35){\line(0,1){10}}
\end{picture}
\caption{Phase diagram for $D=4+1$ NCOS with
\( \tilde{g} N \ll 1 \). \label{figF1D4ph3} }
\end{figure}

For 4+1 dimensional NCOS the curvature 
of the supergravity dual in units of \( \lsb^{-2} \) is \cite{Harmark:2000wv}
\begin{equation}
\mathcal{C} = \frac{b}{\tilde{r}^2} H^{-1/2}
\end{equation}
Thus, for $\tilde{g} N < 1$ we 
need $\tilde{r} \gg \sqrt{b} / ( \tilde{g}N )$
while for $\tilde{g} N > 1$ we need \( \tilde{r} \gg \sqrt{b} \)
in order for \( \mathcal{C} \ll 1 \).

Consider first \( \tilde{g} N \gg 1 \).
When $\tilde{r} \sim L $ we flow from YM to NCOS with the space-time
commutator $[\tilde{t},\tilde{x}^1] = ib$. 
The 4+1 dimensional NCOS theory flows into OM theory when
\( \gsba e^{\phi} \sim 1 \).
This gives two possible phase diagrams, which we have depicted
in the figures \ref{figF1D4ph1} and \ref{figF1D4ph2}.

The case with \( \tilde{g}N \ll 1 \) is depicted in figure \ref{figF1D4ph3}.
This case corresponds to weakly coupled NCOS theory.
The NCOS theory flows to OM theory at \( u \sim N^{1/3}/\lm \).

The thermodynamics of the 4+1 dimensional NCOS theory 
from the supergravity dual
is \cite{Harmark:2000wv}
\begin{equation}
T = \frac{3}{4\pi}\frac{\sqrt{\tilde{r}_0}}{\sqrt{\pi N \tilde{g} b^{3/2}}}
\spa
S = \frac{1}{12 \pi^4} \tilde{V}_4 
\frac{\sqrt{\pi N \tilde{g} b^{3/2}}}{\tilde{g}^2 b^4 }
\tilde{r}_0^{5/2} 
\end{equation}
\begin{equation}
E = \frac{5}{96 \pi^5} \frac{\tilde{V}_4}{\tilde{g}^2 b^4} \tilde{r}_0^3
\spa
F = - \frac{1}{96 \pi^5} \frac{\tilde{V}_4}{\tilde{g}^2 b^4} \tilde{r}_0^3
\end{equation}
\begin{equation}
F= - \frac{2^7 \pi^4}{3^7} N^3 \tilde{g} \sqrt{b} \tilde{V}_4 T^6
\end{equation}
As noted in \cite{Harmark:2000wv} this thermodynamics is equivalent
to that of ordinary 4+1 dimensional YM for large N and strong 't Hooft
coupling.

The thermodynamics of OM theory from its supergravity dual is
\begin{equation}
T = \frac{3}{4\pi}\frac{\sqrt{\tilde{r}_0}}{\sqrt{\pi N \lm^3}}
\spa
S = \frac{1}{24 \pi^5} \tilde{V}_5 
\frac{\sqrt{\pi N \lm^3}}{\lm^9 }
\tilde{r}_0^{5/2} 
\end{equation}
\begin{equation}
E = \frac{5}{192 \pi^6} \frac{\tilde{V}_5}{\lm^9} \tilde{r}_0^3
\spa
F = - \frac{1}{192 \pi^6} \frac{\tilde{V}_5}{\lm^9} \tilde{r}_0^3
\end{equation}
\begin{equation}
F = - \frac{2^6 \pi^3}{3^7} N^3 \tilde{V}_5 T^6
\end{equation}
We see that the thermodynamics of OM theory is equivalent to that
of $(2,0)$ SCFT for large $N$.

\section{$(2,0)$ OBLST and OM theory}
\label{secOBLST20}
\subsection{Introduction to $(2,0)$ OBLST}

The $(2,0)$ OBLST is defined as the D2-NS5 bound-state 
in the decoupling limit
\begin{equation}
\gsna \rightarrow 0 \spa \lsn = \mbox{fixed} \spa 
A_{012} \rightarrow A^{\rm critical}_{012} 
\end{equation}
We show in the following that this limit follows both from
using the near-horizon/decoupling limit of OM theory found
in Section \ref{secOMlimit} and 
from doing a T-duality on the $(1,1)$ OBLST.
The $(2,0)$ OBLST has an LST-string and since 
$(2,0)$ OBLST reduces to OM theory for low energies, it also has
an open membrane.
The T-duality between the two OBLSTs 
is shown to relate the open membrane to the
open string of $(1,1)$ OBLST.

At high energies the LST-string dominates and the thermodynamics
has LST Hagedorn behavior. The R-symmetry is $SO(4)$ for these energies,
but at low energies we get OM theory and the R-symmetry is enhanced
to $SO(5)$. This we show using the supergravity dual of $(2,0)$ OBLST.

The decoupling and near-horizon limits of the D2-NS5 bound state
have previously been studied in \cite{Alishahiha:1999ci,Alishahiha:2000er}.

\subsection{D2-NS5 bound state from M2-M5 on a transverse circle}

The D2-NS5 bound-state in type IIA string theory can be considered as 
an M2-M5 bound-state localized on a transverse circle.
Thus, from the M2-M5 bound state in section \ref{secM2M5F1D4}
we get the metric%
\footnote{We write only the extremal version of this solution so in comparing 
with the non-extremal M2-M5 solution \eqref{M2M5met} and 
\eqref{M2M5pot1}-\eqref{M2M5pot3}
one should use that \( r_0^3 \sinh^2 \alpha = \pi N l_p^3 \)
for \( r_0 \rightarrow 0 \).}
\begin{eqnarray}
\label{M2M5zmet}
ds_{11}^2 &=& (HD)^{-1/3} \Big[ D \Big( - dt^2 +(dx^1)^2 + (dx^2)^2 \Big)
+ (dx^3)^2 +(dx^4)^2 + (dx^5)^2 
\nn \\ &&
+ H \Big( dz^2 + dr^2 + r^2 d\Omega_3^2 \Big) \Big]
\end{eqnarray}
with
\begin{eqnarray}
\label{M2M5zdef}
H &=& 1 + \sum_{n=-\infty}^{\infty} 
\frac{\pi N l_p^3}{(r^2 + (z+2\pi n R_T)^2 )^{3/2}}
\\
D^{-1} &=& \cosh^2 \theta - \sinh^2 \theta H^{-1}
\end{eqnarray}
where $z$ is the coordinate of the transverse circle with 
the asymptotic radius $R_T$.

For \( r \gg R_T \) we have
\begin{equation}
H = 1 + \frac{1}{\pi R_T } \frac{\pi N l_p^3}{r^2}
\end{equation}

We now consider the \( z \) coordinate as the eleven dimensional
coordinate. Thus by the usual M/IIA correspondence we
have \( R_T = \gsna \lsn \) and \( l_p^3 = R_T \lsn^2 \).
The D2-NS5 bound-state has the string-frame
metric $ds_{10}^2$ and the dilaton $e^\phi$ given by the formula
\begin{equation}
ds_{11}^2 = e^{-\frac{2}{3} \phi} ds_{10}^2 + e^{\frac{4}{3}\phi} dz^2
\end{equation}
This gives the D2-NS5 solution
\begin{eqnarray}
ds_{10}^2 &=& D^{-1/2} \Big[ D \Big( - dt^2 +(dx^1)^2 + (dx^2)^2 \Big)
+ (dx^3)^2 +(dx^4)^2 + (dx^5)^2 
\nn \\ &&
+ H \Big( dr^2 + r^2 d\Omega_3^2 \Big) \Big]
\end{eqnarray}
\begin{equation}
e^{2\phi} = H D^{-1/2}
\end{equation}
\begin{eqnarray}
A_{012} &=& - \sin \hat{\theta} D H^{-1} 
\\
A_{345} &=& \tan \hat{\theta} H^{-1}
\end{eqnarray}

\subsection{Supergravity dual of $(2,0)$ OBLST}

The near-horizon limit of the solution \eqref{M2M5zmet} is
\begin{equation}
\label{M2M5zlim1}
l_p \rightarrow 0 \spa
\tilde{r} = \frac{\lm^3}{l_p^3} r \spa
\tilde{z} = \frac{\lm^3}{l_p^3} z \spa
\lm^6 = l_p^6 \cosh \theta
\end{equation}
\begin{equation}
\label{M2M5zlim2}
\tilde{x}^i = \frac{l_p^3}{\lm^3} x^i, \ \ i=0,1,2 \spa
\tilde{x}^j = \frac{\lm^3}{l_p^3} x^j,\ \ j=3,4,5 
\end{equation}
This we obtained from the OM near-horizon limit 
\eqref{M2M5lim1}-\eqref{M2M5lim2} since the $(2,0)$ OBLST limit
should be consistent with the OM theory limit.

The eleven dimensional supergravity 
dual of $(2,0)$ OBLST is therefore given by%
\footnote{The construction of the
supergravity dual of $(2,0)$ OBLST presented here 
is similar to that of the D2-NS5 near-horizon solution presented in 
\cite{Alishahiha:2000er}. The D2-NS5 near-horizon solution of 
\cite{Alishahiha:2000er} describes $(2,0)$ OBLST on a magnetic circle.
We thank M. Alishahiha for a discussion about this point.}
\begin{eqnarray}
\label{NHOB20met}
ds_{11}^2 &=& \frac{l_p^2}{\lm^2} H^{-2/3} J^{1/3} \left[
H  J^{-1} \Big( - d\tilde{t}^2 
+ (d\tilde{x}^1)^2 + (d\tilde{x}^2)^2 \Big)
\right. \nn \\ && \left.
+ (d\tilde{x}^3)^2 + (d\tilde{x}^4)^2 + (d\tilde{x}^5)^2 
+ H \Big( d\tilde{z}^2 + d\tilde{r}^2 + \tilde{r}^2 d\Omega_3^2 \Big) \right]
\end{eqnarray}
\begin{equation}
\label{NHOB20pot}
C_{012} = - \frac{l_p^3}{\lm^3} J^{-1} \spa
C_{345} = \frac{l_p^3}{\lm^3} H^{-1}
\end{equation}
\begin{equation}
J = \sum_{n=-\infty}^{\infty} 
\frac{\pi N \lm^3}{(\tilde{r}^2 
+ (\tilde{z}+2\pi n \tilde{R}_T)^2 )^{3/2}}
\spa
H = 1 + J
\end{equation}

The limit \eqref{M2M5zlim1}-\eqref{M2M5zlim2} is
precisely consistent with our requirements of a limit of 
$(2,0)$ OBLST. 
We need that $r/R_T$ is finite in the near-horizon limit 
so since \( l_p^3 = \gsna \lsn^3 \) and \( R_T = \gsna \lsn \)
we see that the OBLST limit is \( g_s \rightarrow 0 \)
and \( \lsn \) kept fixed. But this is the limit of ordinary LST
so we should have closed strings of tension \( 1/ (2\pi \lsn^2) \)
in $(2,0)$ OBLST. Note that $\tilde{R}_T = \lm^3 / \lsn^2$.

We observe that the supergravity dual of $(2,0)$ OBLST given by 
\eqref{NHOB20met}-\eqref{NHOB20pot}
reduces to the supergravity dual of $(2,0)$ LST given in 
\cite{Itzhaki:1998dd,Aharony:1998ub} when $\tilde{r} \ll N^{1/3} \lm$. 

For \( \tilde{r} \ll \tilde{R}_T \) we should 
consider \eqref{NHOB20met}-\eqref{NHOB20pot}
as an approximate description of $(2,0)$ OBLST which 
for small $\tilde{r}$ continuously should flow to the OM theory 
supergravity dual given by \eqref{NHM2M5met}-\eqref{NHM2M5pot}.

When \( \tilde{r} \gg \tilde{R}_T \) and \( \gsna e^{\phi} \ll 1 \)
we can use a ten-dimensional description
via weakly coupled type IIA string theory.
The limit \eqref{M2M5zlim1}-\eqref{M2M5zlim2} translates into the
limit
\begin{equation}
\label{D2NS5lim1}
\gsna \rightarrow 0 \spa
\lsn = \mbox{fixed} \spa
\tilde{r} = \frac{\lm^3}{\gsna \lsn^3} r \spa
\lm^6 = \gsna^2 \lsn^6 \cosh \theta
\end{equation}
\begin{equation}
\label{D2NS5lim2}
\tilde{x}^i = \frac{\gsna \lsn^3}{\lm^3} x^i, \ \ i=0,1,2 \spa
\tilde{x}^j = \frac{\lm^3}{\gsna \lsn^3} x^j,\ \ j=3,4,5 
\end{equation}
The type IIA near-horizon solution is then
\begin{eqnarray}
ds_{10}^2 &=& H^{-1/2} \frac{L}{\tilde{r}} \left[
H \frac{\tilde{r}^2}{L^2} \Big( -d\tilde{t}^2 
+ (d\tilde{x}^1)^2 + (d\tilde{x}^2)^2 \Big)
\right. \nn \\ && \left.
+ (d\tilde{x}^3)^2 + (d\tilde{x}^4)^2 + (d\tilde{x}^5)^2 
+ H \Big( d\tilde{r}^2 + \tilde{r}^2 d\Omega_3^2 \Big) \right]
\end{eqnarray}
\begin{equation}
\gsna^2 e^{2\phi} = \frac{\lm^6}{\lsn^6} H^{1/2} \frac{L}{\tilde{r}}
\end{equation}
\begin{equation}
A_{012} = - \frac{\gsna \lsn^3}{\lm^3} \frac{\tilde{r}^2}{L^2}
\spa
A_{345} = \frac{\gsna \lsn^3}{\lm^3} H^{-1}
\end{equation}
\begin{equation}
L^2 = \lsn^2 N \spa
H = 1 + \frac{L^2}{\tilde{r}^2}
\end{equation}
%

\subsection{T-duality on an electric circle: From open membranes 
to open strings}
\label{secTdual}

Since we believe that $(2,0)$ OBLST has an open membrane of tension
$1/\lm^3$ and that $(1,1)$ OBLST has an open string of tension
$1/b$ it is natural to ask whether this is consistent with T-duality.
From point of view of the bulk, T-duality on an electric circle 
in $(2,0)$ OBLST would give $(1,1)$ OBLST, since it takes
D2-NS5 into D1-NS5. 
In this section we test that this is also consistent with the 
decoupling limits. 
In section \ref{secchain} we develop this further and connect
5 different bound states and their decoupling limits
in a duality-chain.

We take $x^2$ as the coordinate of the electric circle with 
radius $R_E$. 
From \eqref{M2M5zlim1}-\eqref{M2M5zlim2} and 
\eqref{D2NS5lim1}-\eqref{D2NS5lim2} we get
\begin{equation}
\tilde{R}_E = \frac{\gsna \lsn^3}{\lm^3} R_E
\spa
\tilde{R}_T = \frac{\lm^3}{\gsna \lsn^3} R_T
\end{equation}
Since \( R_T = \gsna \lsn \) we have
\begin{equation}
\tilde{R}_T = \frac{\lm^3}{\lsn^2}
\end{equation}
A T-duality in $x^2$ gives
\begin{equation}
\label{RErel}
\gsnb = \gsna \frac{\lsn}{R_E} = \frac{R_T}{R_E} 
= \frac{\gsna^2 \lsn^4}{\lm^3} \frac{1}{\tilde{R}_E}
\end{equation}
Since \( l_s \) is fixed in both $(1,1)$ and $(2,0)$ OBLST
this means that \( \gsnb \propto \gsna^2 \). By comparing
\eqref{D1NS5lim1}-\eqref{D1NS5lim2} with
\eqref{D2NS5lim1}-\eqref{D2NS5lim2} 
we see that this is exactly what we need for the decoupling/near-horizon
limits of $(2,0)$ and $(1,1)$ OBLST to be consistent with each other. 
Moreover, we see that we need
\begin{equation}
\label{comprel}
\frac{\lm^3}{\gsna \lsn^3} = \frac{\sqrt{b}}{\sqrt{\gsnb} \lsn}
\end{equation}
Using \eqref{RErel} and \eqref{comprel} we get that
\begin{equation}
\tilde{R}_E = \frac{\gsna^2}{\gsnb} \frac{\lsn^4}{\lm^3}
= \frac{\lm^3}{b}
\end{equation}
Thus, the T-duality between the decoupling limits of $(1,1)$ and $(2,0)$
OBLST requires that
\begin{equation}
\frac{1}{b} = \frac{\tilde{R}_E}{\lm^3}
\end{equation}
This relation means that the open string of $(1,1)$ OBLST with
tension $1/b$ is the open membrane of $(2,0)$ OBLST wrapped
around the electric circle of radius $\tilde{R}_E$. 
Thus, the open membrane in $(2,0)$ OBLST
and the open string in $(1,1)$ OBLST are related by T-duality.

We elaborate further on this in Section \ref{secchain}.

\subsection{Phase structure and thermodynamics}

As already mentioned, the $(2,0)$ OBLST has both the open membrane
with tension $1/\lm^3$ as in OM theory, 
and also the LST-string with tension $1/(2\pi \lsn^2)$.
We now consider the phase structure of $(2,0)$ OBLST.
We parameterize the phase diagrams with the rescaled radial
coordinate $\tilde{r}$. This is not an energy coordinate,
but any energy coordinate should be increasing with $\tilde{r}$
and we can therefore use it to find the succession 
of transition points.

We consider two possible phase diagrams depicted in figure 
\ref{figOB20a} and \ref{figOB20b}.
For both diagrams we have that at $\tilde{r} \sim \tilde{R}_T$
we have a transition point where for lower energies we have
$SO(5)$ R-symmetry and for higher energies $SO(4)$ R-symmetry.
In the supergravity solution this can be understod from
the observation that at $\tilde{r} \sim \tilde{R}_T$ the radius
of the $S^3$ in the metric \eqref{NHOB20met} is of the same
order as the radius $\tilde{R}_T$ of the transverse circle.
Thus, at $\tilde{r} \ll \tilde{R}_T$ the supergravity dual of $(2,0)$
OBLST is in fact the supergravity dual of OM theory, given in
Section \ref{secOMlimit}.

The curvature in units of $\lsn^{-2}$ 
of the supergravity dual for $\tilde{r} \gg \tilde{R}_T$ is
\begin{equation}
\mathcal{C} = \frac{1}{N} \frac{1}{\sqrt{1+ \frac{\tilde{r}^2}{\lsn^2 N}}}
\end{equation}
For $\tilde{r} \ll \tilde{R}_T$ the curvature is given by \eqref{OMcurv}.
In the following we work with $N \gg 1$.

We first consider the phase diagram of figure \ref{figOB20a}.
For low energies we have $(2,0)$ SCFT with $SO(5)$ as the R-symmetry
group. This is described by the ${\rm AdS}_7 \times S^4$ supergravity dual.
From Section \ref{secOMphases} we know that at $\tilde{r} \sim N^{1/3} \lm$
we have OM theory, described by delocalized M2-branes.
At $\tilde{r} \sim \tilde{R}_T$ 
the R-symmetry is broken to $SO(4)$ and we go into $(2,0)$ OBLST.
However, we do not enter the weakly coupled type IIA description
before the transition point $\gsna e^{\phi} \sim 1$,
which is at $\tilde{r} \sim \sqrt{N} \lm^6 / \lsn^5$.
Thus, we have either delocalized M2-branes or delocalized D2-branes
describing the $(2,0)$ OBLST phase.

In the second phase diagram, depicted in figure \ref{figOB20b},
we also start at low energies with $(2,0)$ SCFT.
We then proceed to the ordinary $(2,0)$ LST at 
$\tilde{r} \sim \tilde{R}_T$. At $\tilde{r} \sim \sqrt{N} \lm^3 / \lsn^2$
we enter the weakly coupled type IIA description so that 
the $(2,0)$ LST is described by NS5-branes.
At $\tilde{r} \sim \sqrt{N} \lsn$ we enter the $(2,0)$ OBLST phase
with a non-commutative space-time. 

\begin{figure}[h]
\begin{picture}(390,75)(0,0)
\put(5,40){\vector(1,0){380}}
\put(382,50){$\tilde{r}$}
\put(20,5){\shortstack{$(2,0)$ SCFT \\ ${\rm AdS}_7 \times S^4$}}
\put(115,5){\shortstack{OM theory \\ M2-branes}}
\put(205,5){\shortstack{$(2,0)$ OBLST \\ M2-branes}}
\put(300,5){\shortstack{$(2,0)$ OBLST \\ D2-branes}}
\put(2,50){0}
\put(5,35){\line(0,1){10}}
\put(80,55){$N^{1/3} \lm $}
\put(96,35){\line(0,1){10}}
\put(186,55){$\tilde{R}_T$}
\put(192,35){\line(0,1){10}}
\put(275,55){$\frac{\sqrt{N} \lm^6}{\lsn^5}$}
\put(288,35){\line(0,1){10}}
\end{picture}
\caption{Phase diagram for $(2,0)$ OBLST. \label{figOB20a} }
\end{figure}

\begin{figure}[h]
\begin{picture}(390,75)(0,0)
\put(5,40){\vector(1,0){380}}
\put(382,50){$\tilde{r}$}
\put(20,5){\shortstack{$(2,0)$ SCFT \\ ${\rm AdS}_7 \times S^4$}}
\put(115,5){\shortstack{$(2,0)$ LST \\ M5-branes}}
\put(209,5){\shortstack{$(2,0)$ LST \\ NS5-branes}}
\put(300,5){\shortstack{$(2,0)$ OBLST \\ D2-branes}}
\put(2,50){0}
\put(5,35){\line(0,1){10}}
\put(90,55){$\tilde{R}_T$}
\put(96,35){\line(0,1){10}}
\put(179,55){$\frac{\sqrt{N} \lm^3}{\lsn^2}$}
\put(192,35){\line(0,1){10}}
\put(275,55){$\lsn \sqrt{N}$}
\put(288,35){\line(0,1){10}}
\end{picture}
\caption{Phase diagram for $(2,0)$ OBLST. \label{figOB20b} }
\end{figure}

From a non-extremal version of the metric \eqref{NHOB20met} 
the thermodynamics of $(2,0)$ OBLST for $\tilde{r} \gg \tilde{R}_T$ is
found to be
\begin{equation}
\label{OB20therm1}
T = \frac{1}{2\pi \lsn \sqrt{N}}
\spa
S = \sqrt{N} \frac{\tilde{V}_5 }{(2\pi)^4} \frac{1}{\lsn \lm^6} \tilde{r}_0^2
\end{equation}
\begin{equation}
\label{OB20therm2}
E = \frac{\tilde{V}_5 }{(2\pi)^5} \frac{1}{\lsn^2 \lm^6} \tilde{r}_0^2
\spa
F = 0
\end{equation}
The Hagedorn temperature of $(2,0)$ OBLST is
\begin{equation}
\tlst = \frac{1}{2\pi \lsn \sqrt{N}}
\end{equation}
Thus, we see that for $\tilde{r} \gg \tilde{R}_T$ we have $T \sim \tlst$
so that the LST-strings dominate for $\tilde{r} \gg \tilde{R}_T$.
As discussed in \cite{Maldacena:1996ya,Maldacena:1997cg,Harmark:2000hw} 
the thermodynamics 
\eqref{OB20therm1}-\eqref{OB20therm2} exhibits leading order Hagedorn
behavior, and one can calculate \cite{Correia:2000} 
that since the supergravity dual
consist of delocalized D2-branes in the UV-region, the 
entropy has the critical behavior $S(T) \propto (\tlst-T)^{-2/3}$,
just as for $(1,1)$ OBLST. Thus, the critical behavior
of the entropy for high energies
in $(2,0)$ OBLST is different from that in $(2,0)$ LST.

From comparing the phases and thermodynamics of $(1,1)$ OBLST and
$(2,0)$ we see that there are many similarities, as one would expect
from T-dual theories. 
The LST-strings dominate the thermodynamics for high energies
in both cases, and the open string and open membrane only appears
as phases in the phase diagrams when they are sufficiently light.

\subsection{Branes in $(2,0)$ OBLST}

In the ordinary $(2,0)$ LST we have besides the LST-string with
tension $1/(2\pi \lsn^2)$ the d1, d3 and d5 branes\cite{Losev:1997hx}.
These origins from having open D2, D4 and D6 branes
stretching between NS5-branes.

The $(2,0)$ OBLST still have the LST-string, but the
D-membrane stretching between two D2-NS5 bound-states induces
an open membrane which gives OM theory at low energy. 
At low energies this open membrane reduces to a d1-brane
in $(2,0)$ LST, or to a tensionless string in $(2,0)$ SCFT.
Using similar arguments as for $(1,1)$ OBLST, we
expect that the d3 and d5-branes are part of $(2,0)$ OBLST.

Thus, we see that the brane-spectra of $(1,1)$ and $(2,0)$ OBLST
are related by T-duality. 

\section{A duality-chain of theories}
\label{secchain}

\begin{figure}[h]
\begin{picture}(390,290)(0,0)
\put(125,215){\framebox(140,70)[l]{}}
\put(15,110){\framebox(140,70)[l]{}}
\put(235,110){\framebox(140,70)[l]{}}
\put(15,5){\framebox(140,70)[l]{}}
\put(235,5){\framebox(140,70)[l]{}}
\put(85,75){\vector(0,1){35}}
\put(85,110){\vector(0,-1){35}}
\put(305,75){\vector(0,1){35}}
\put(305,110){\vector(0,-1){35}}
\put(155,40){\vector(1,0){80}}
\put(235,40){\vector(-1,0){80}}
\put(140,180){\vector(0,1){35}}
\put(140,215){\vector(0,-1){35}}
\put(250,180){\vector(0,1){35}}
\put(250,215){\vector(0,-1){35}}
\put(135,240){\shortstack[l]{M2-M5 bound state\\ M-theory: $l_p$ \\  
OM theory: $\lm$}}
\put(23,130){\shortstack[l]{F1-D4 bound state\\ Type IIA: $\gsba,\lsb$ \\  
$D=4+1$ NCOS: $b,\tilde{g}_4$}}
\put(23,17){\shortstack[l]{F1-D5 bound state\\ Type IIB: $\gsbb,\lsb$ \\  
$D=5+1$ NCOS: $b,\tilde{g}_5$ \\ $(1,1)$ OBLST: $b,\lsn$}}
\put(240,17){\shortstack[l]{D1-NS5 bound state\\ Type IIB: $\gsnb,\lsn$ \\  
$D=5+1$ NCOS: $b,\tilde{g}_5$ \\ $(1,1)$ OBLST: $b,\lsn$}}
\put(240,130){\shortstack[l]{D2-NS5 bound state\\ Type IIA: $\gsna,\lsn$ \\  
 $(2,0)$ OBLST: $\lm,\lsn$}}
\put(150,195){\shortstack[l]{$S_A(E)$}}
\put(95,90){\shortstack[l]{$T(T)$}}
\put(260,195){\shortstack[l]{$S_A(T)$}}
\put(315,90){\shortstack[l]{$T(E)$}}
\put(185,45){\shortstack[l]{$S_B$}}
\end{picture}
\caption{The chain of theories and bound-states 
related by S- and T-dualities. $S_A(E)$ and $S_A(T)$ means the
type IIA S-duality in the electrical and transverse direction, respectively.
$T(E)$ and $T(T)$ means a T-duality in the electrical and transverse
direction, respectively. $S_B$ means type IIB S-duality.
\label{figchain} }
\end{figure}
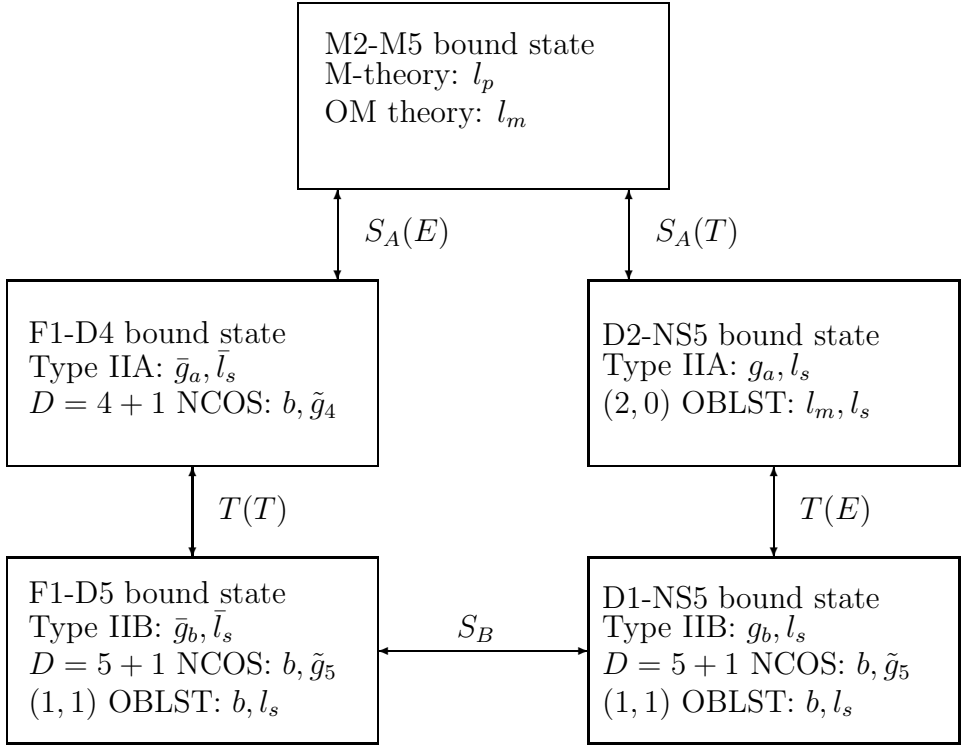

In this section%
\footnote{The content of this section was developed 
in collaboration with N. A. Obers.}
we systematically explore how the T-dualities and
S-dualities connects the various bound-states and their
decoupling limits that we have been discussing in 
Section \ref{secOBLST11}-\ref{secOBLST20}, 
in order test the consistency of these limits and also to relate
the parameters of the theories.
Some of the discussion has already appeared in earlier sections,
but we repeat it here for clarity.
The duality-chain is depicted in Figure \ref{figchain}.

All of the bound-states in the chain can be seen as the M2-M5 bound-state
on an electric circle and a transverse circle. 
In eleven dimensions the M2-M5 bound-state on a transverse
circle is given by \eqref{M2M5zmet}-\eqref{M2M5zdef}.
We take \( x^2 \) to be the coordinate for the electric circle of 
radius $R_E$ and $z$ to be the coordinate for the transverse circle
with radius $R_T$.

For all the various decoupling limits we have
\begin{equation}
\label{Alllim1}
\cosh \theta \rightarrow \infty \spa
\tilde{r} = \sqrt{\cosh \theta} r \spa
\tilde{z} = \sqrt{\cosh \theta} z 
\end{equation}
\begin{equation}
\label{Alllim2}
\tilde{x}^i = \frac{1}{\sqrt{\cosh \theta}} x^i, \ \ i=0,1,2 \spa
\tilde{x}^j = \sqrt{\cosh \theta} x^j,\ \ j=3,4,5 
\end{equation}
which gives
\begin{equation}
\tilde{R}_E = \frac{1}{\sqrt{\cosh \theta}} R_E \spa
\tilde{R}_T = \sqrt{\cosh \theta} R_T 
\end{equation}
Thus, all the decoupling limits are specified by the relation between
\( \cosh \theta \) and the parameters of M/string theory, and
the parameters of the world-volume theories.

Thus, starting from the top with the M2-M5 bound-state, we have
\begin{equation}
\cosh \theta = \frac{\lm^6}{l_p^6}
\end{equation}
where \( 1/\lm^3 \) is the tension of the open membrane in OM theory
and $(2,0)$ OBLST.

Choosing $x^2$ as the eleventh direction
we go to the F1-D4 bound-state and we have
\begin{equation}
R_E = \gsba \lsb \spa
l_p^3 = \gsba \lsb^3 \spa
\cosh \theta = \frac{b}{\lsb^2} \spa
\tilde{g}_4 = \frac{\gsba \lsb^2}{b}
\end{equation}
with the NCOS open string coupling \( \tilde{g}_4 \) and tension $1/b$.
This gives the relations
\begin{equation}
\tilde{R}_E = \tilde{g}_4 \sqrt{b} \spa
\frac{1}{b} = \frac{\tilde{R}_E}{\lm^3}
\end{equation}
Thus, as already mentioned in Section \ref{secOMlimit}, this is interpreted
\cite{Gopakumar:2000ep,Bergshoeff:2000ai}
as the fact that the NCOS open string in 4+1 dimensions is an
 open membrane in 5+1 dimensions wrapped on the electric circle of 
radius $\tilde{R}_E$, and for strong coupling
the electric circle is large and we flow into decompactified 
OM theory.

Making a T-duality in the transverse direction $z$ we go to the
F1-D5 bound-state. We have
\begin{equation}
\gsbb = \gsba \frac{\lsb}{R_T} = \frac{R_E}{R_T} \spa
\cosh \theta = \frac{b}{\lsb^2} \spa 
\tilde{g}_5 = \frac{\gsbb \lsb^2}{b}
\end{equation}
where \( \tilde{g}_5 \) is the 5+1 dimensional NCOS open string coupling.
We also define the T-dual radius \( R'_T = \lsb^2 / R_T \) along with
its rescaled version \( \tilde{R}'_T \).
This gives 
\begin{equation}
\label{NCOSTdual}
\tilde{g}_5 = \tilde{g}_4 \frac{\sqrt{b}}{\tilde{R}_T} \spa
\tilde{R}'_T \tilde{R}_T = b
\end{equation}
We see that this can be interpreted as a NCOS T-duality (similar
interpretations for other cases have been done in 
\cite{Bergshoeff:2000ai,Kawano:2000gn}).
Thus, the bulk T-duality on the F1-D$p$ bound states induces 
a world-volume T-duality in the NCOS theories relating the
T-dual string couplings and radii by the NCOS string tension $1/b$.

The type IIB S-duality takes us into the D1-NS5 bound-state
for which we have
\begin{equation}
\label{D1NS5results}
\gsnb = \frac{1}{\gsbb} \spa
\lsn^2 = \gsbb \lsb^2 \spa
\cosh \theta = \frac{b}{\gsnb \lsn^2} \spa
\tilde{g}_5 = \frac{\lsn^2}{b}
\end{equation}
as explained in Section \ref{secSupOB11}.

If again start from the top and instead choose $z$ as the eleventh direction 
we go from the M2-M5 to the D2-NS5 bound-state. We have
\begin{equation}
R_T = \gsna \lsn \spa
l_p^3 = \gsna \lsn^3 \spa
\cosh \theta = \frac{\lm^6}{\gsna^2 \lsn^6}
\end{equation}
Here the world-volume theory is $(2,0)$ OBLST with the parameters
$\lsn$, $\lm$ and $\tilde{R}_E$.

Making a T-duality in the $x^2$ direction we obtain the D1-NS5 
bound-state. As already explained in Section \ref{secTdual}, we have
\begin{equation}
\gsnb = \gsna \frac{\lsn}{R_E} = \frac{R_T}{R_E} \spa
\cosh \theta = \frac{\lm^3}{\gsnb \lsn^2 \tilde{R}_E }
\end{equation}
We again note that $\cosh \theta$ exactly has the right dependence
on the string coupling $\gsnb$ that makes us able to compare this
T-dualized limit to the limit obtained in \eqref{D1NS5results}
by going the other way in the chain.
We now define the T-dual radius \( R'_E = \lsn^2 / R_E \) 
of the electric circle. Since \( x^2 \) after the T-duality
is a magnetic coordinate, $R'_E$ scales oppositely to $R_E$.
Using this and comparing with \eqref{D1NS5results} we therefore get
\begin{equation}
\frac{1}{b} = \frac{\tilde{R}_E}{\lm^3} \spa
\tilde{R}'_E \tilde{R}_E = \lsn^2
\end{equation}
Thus, we see that the T-duality in the bulk now induces a 
little-closed-string-T-duality in the OBLST, since contrary to
\eqref{NCOSTdual} the T-duality is in terms of the 
length scale $l_s$ of the closed strings in OBLST.
One could speculate that this means that open strings of tension
$1/(2\pi \lsn^2)$ can end on the open string/membrane
in OBLST. 
Certainly, this would make sense from the bulk point of view, since
the open string/membrane origins from D1 or D2-branes stretching
between the D1-NS5 or D2-NS5 bound-states.

\section*{Acknowledgments}

We thank N.A. Obers for early participation and useful discussions
and M. Alishahiha, D. Berman, E. Cheung, J. Correia, S. Minwalla, J. Nielsen 
and J. L. Petersen for useful discussions.

\ \newline {\bf Note added:} \newline
While writing this paper, we received the paper \cite{Berman:2000zw}
which also considers a supergravity dual of OM theory
using a different approach than the one we have in
section \ref{secOMdual}.

In the final stages of writing this paper we were 
made aware that N. Seiberg and A. Strominger have presented
similar ideas as of this paper in their talks at Strings 2000, july 10-15.
We were subsequently informed that these ideas will appear in
a revised version of the paper \cite{Gopakumar:2000ep} 
by R. Gopakumar, S. Minwalla, N. Seiberg and A. Strominger.


\addcontentsline{toc}{section}{References}


\providecommand{\href}[2]{#2}\begingroup\raggedright\endgroup

\end{document}